# Anisotropic magnetoresistance in antiferromagnetic $Sr_2IrO_4$

C. Wang[1,2], H. Seinige[1,2], G. Cao[3], J.-S. Zhou[2], J. B. Goodenough[2], M. Tsoi[1,2]

[1]*Physics Department, University of Texas at Austin, Austin, Texas 78712, USA*

[2]*Texas Materials Institute, University of Texas at Austin, Austin, Texas 78712, USA*

[3]*Center for Advanced Materials, Department of Physics and Astronomy, University of Kentucky, Lexington, Kentucky 40506, USA*

We report point-contact measurements of anisotropic magnetoresistance (AMR) in a single crystal of antiferromagnetic (AFM) Mott insulator $Sr_2IrO_4$. The point-contact technique is used here as a local probe of magnetotransport properties on the nanoscale. The measurements at liquid nitrogen temperature revealed negative magnetoresistances (MRs) (up to 28%) for modest magnetic fields (250 mT) applied within the $IrO_2$ a-b plane and electric currents flowing perpendicular to the plane. The angular dependence of MR shows a crossover from four-fold to two-fold symmetry in response to an increasing magnetic field with angular variations in resistance from 1-14%. We tentatively attribute the four-fold symmetry to the crystalline component of AMR and the field-induced transition to the effects of applied field on the canting of AFM-coupled moments in $Sr_2IrO_4$. The observed AMR is very large compared to the crystalline AMRs in 3d transition metal alloys/oxides (0.1-0.5%) and can be associated with the large spin-orbit interactions in this 5d oxide while the transition provides evidence of correlations between electronic transport, magnetic order and orbital states. The finding of this work opens an entirely new avenue to not only gain a new insight into physics associated with spin-orbit coupling but also better harness the power of spintronics in a more technically favorable fashion.

Antiferromagnetic (AFM) spintronics [1-4] is a new emerging field of material science and device physics aiming to explore unique properties of AFMs and implementing them as active ingredients in spintronic applications. AFM materials share a number of useful functionalities with ferromagnets (FM) such as spin-transfer torque [1, 2], which is predicted to be even stronger in AFMs [1], and exhibit unique interactions with FMs as in the well-known effect of exchange bias [5]. Moreover, they are instrumental in minimizing the cross-talk between nanodevices as AFMs do not produce stray magnetic fields. One of the milestones of AFM spintronics is finding an efficient method for monitoring the magnetic order parameter in AFM materials. Anisotropic magnetoresistance (AMR) [6] and tunneling AMR [7, 8] observed in AFMs are among very promising candidates for this purpose. The canted AFM iridate $Sr_2IrO_4$ (see Fig. S1 in Supplemental Material) is a particularly interesting material for such AMR studies. The strong spin-orbit interaction in this and other iridates drives many fascinating phenomena including the $J_{eff} =1/2$ Mott state [9], possible superconductivity [10], topological insulator [11] and spin liquid [12] behaviors, which makes them an attractive playground for studying physics driven by spin-orbit interactions. As AMR is known to be closely associated with spin-orbit interaction, the strong spin-orbit interaction in this 5d transition metal oxide may favor stronger AMR compared to 3d transition metal alloys and oxides. The recent magnetotransport studies in $Sr_2IrO_4$ single crystals [13, 14] and thin films [6] revealed largely unexplored correlations between electronic transport, magnetic order and orbital states.

Here we present the first observation of the point-contact AMR in single crystals of the AFM Mott insulator $Sr_2IrO_4$ [13-15], which can potentially be used to sense the AFM order parameter in spintronic nanodevices. The point-contact technique allows to probe very small volumes and, therefore, measures electronic transport on a microscopic scale. Point-contact measurements with single crystals of $Sr_2IrO_4$ were intended to examine if the additional local resistance associated with a small contact area between a sharpened Cu tip and the antiferromagnet shows a magnetoresistance (MR) like that seen in bulk crystals. The measurements at liquid nitrogen temperature revealed large MRs (up to 28%) for modest magnetic fields (250 mT) applied within the IrO$_2$ a-b plane. The angular dependence of MR revealed an AMR with an intriguing transition from four-fold to two-fold symmetry in response to an increasing magnetic field. We tentatively attribute the four-fold symmetry to the crystalline component of AMR and the field-induced transition to the effects of applied field on the canting of AFM-coupled moments in Sr$_2$IrO$_4$. These findings open an entirely new avenue to not only gain a quantum-mechanical insight into the new physics but also better harness the power of spintronics in a more technically favorable fashion.

Our sample is a single crystal of Sr$_2$IrO$_4$ (1.5 mm×1 mm×0.5 mm) synthesized via a self-flux technique [16]. The insert to Fig. 1a shows a schematic of our experiment: a point contact between a sharpened Cu tip and the single crystal (001) surface was made with a mechanically controlled differential-screw system described elsewhere [17]; an electrical current is injected through the point contact into the crystal and flows (primarily) along the [001] c-axis into a macroscopic Cu electrode on the back side of the crystal. The system enables us to produce point contacts on the sample's surface with a dimension $a$ from microns down to a few nanometers [18]. The point-contact current-voltage characteristics exhibit largely ohmic behavior, as shown in Supplemental Material Fig. S2. The contact size $a$ can be estimated from the measured contact resistance $R$ with a simple model [18] for diffusive point contacts that gives $R=\rho/2a$, where $\rho$ is the resistivity of Sr$_2$IrO$_4$. Note that the resistive contributions of the Cu tip and Cu back electrode (see insert to Fig. 1a) are negligible due to a much higher conductivity of Cu. Assuming $\rho$ of Sr$_2$IrO$_4$ ≈50 Ωcm at liquid nitrogen temperature [13], this analysis yields $a$ ranging from 45 nm – 4.2 μm for $R$ = 15 kΩ - 1.4 MΩ. In the following, we present results from two representative point contacts with $a \approx$ 3 μm (PC1) and 1 μm (PC2). The results of MR measurements were found reproducible for a dozen contacts with intermediate sizes (see Fig. S3 in Supplemental Material). At liquid nitrogen temperature (77K) we have measured the point-contact magnetoresistance $R(H)$ for different orientations of the applied magnetic field $H$ in the (001) a-b plane of Sr$_2$IrO$_4$ crystal.

Figure 1a shows the point-contact MR – the $R(H)$ dependences – for PC1 (bottom) and PC2

(top). The black (grey) traces are the up- (down-) sweeps from high negative (positive) to high positive (negative) fields. We have measured $R(H)$ traces for different angles $\theta$ between the applied magnetic field $H$ and the [100] a-axis (not shown). All MR traces exhibit negative magnetoresistances with generally similar shapes (like in Fig. 1a) and MR ratios $(R_{max}-R_{min})/R_{min}$ from 6.7-8.4 % for PC1 and 20-28 % for PC2. Interestingly, we find an anisotropy in the point-contact resistance $R$ for different orientations of $H$ (different $\theta$s). This anisotropic magnetoresistance (AMR) is presented in Fig. 1b, which shows $R(\theta)$ for a series of applied fields $H$ obtained from $R(H)$ data as in Fig. 1a. The curves were shifted vertically for clarity. The data shows that $R(\theta)$ has a predominantly four-fold symmetry at low fields (top trace in Fig. 1b; also see grey trace in Fig. 2b) and a two-fold symmetry at high fields (bottom trace in Fig. 1b). This major observation is further illustrated by polar plots in Fig. 1c, which show the four- and two-fold symmetries of the normalized AMR $[R(\theta)-R_{min}]/[R_{max}-R_{min}]$ at $\mu_0 H$ = 40 and 270 mT, respectively. $R(\theta)$ traces at intermediate fields (progressively from top to bottom trace in Fig. 1b = increasing $H$) show a gradual transition from the four- to two-fold symmetry with increasing magnetic field: two out of four 'peaks' in $R(\theta)$ start to shrink and the positions of the remaining two 'peaks' shift by as much as 40 degrees as the applied $H$ increases above the critical field ($\mu_0 H_c \approx$ 200 mT) of Sr$_2$IrO$_4$ meta-magnetic transition (see Fig. S4 in Supplemental Material). Such a field-dependent AMR transition was found to be reproducible in all of our point contacts. Figures 3a and 3b show details of the symmetry transition in a 2D grey-scale plot representation (lighter color indicates higher resistance) for PC1 and PC2, respectively. Both grey-density plots show qualitatively similar transitions in AMR from four- (at low fields) to two-fold symmetry (at higher fields). For the smaller contact (PC2), the symmetry transition occurs at a somewhat larger field (~ 60 mT) compared to that for the larger PC1 contact (~ 40 mT).

To further quantify the data in Fig. 1 we note that the amplitude of $R(\theta)$ variations – the magnitude of AMR=$[R(\theta)-R_{min}]/R_{min}$ – changes with increasing $H$. The four-fold variations in $R(\theta)$ at low fields ($\mu_0 H$ < 70 mT) are much smaller than the two-fold variations at higher fields (see different traces in Fig. 1b). Figure 2a shows that the magnitude of AMR is a non-monotonic function of $H$ and peaks at around 120 mT. Finally the 'coercive' field $H^*$ where $R(H)$ has a maximum exhibits small variations as a function of $\theta$. Figure 2b shows that these variations in $H^*(\theta)$ are somewhat correlated with the four-fold variations in $R(\theta)$ at low fields (grey trace) but have a predominantly two-fold character.

We would like to point out that the AMR observed in our experiments cannot be explained by the conventional AMR in polycrystalline magnetic conductors defined solely by the relative angle between the current direction and magnetic moments [19]. The current is being injected vertically through the point contact (perpendicular to the a-b plane; see experimental setup in inset to Fig. 1a) and is not expected to see any changes in its relative orientation with respect to the magnetic moments of $Sr_2IrO_4$ as the applied magnetic field is rotated within the sample's basal a-b plane. Instead, it is the relative angle between the moments and the crystal axes that may change as the field rotates. Therefore the observed anisotropy in resistance can be tentatively attributed to the *crystalline component* of AMR. This spin-orbit (SO) coupling induced effect arises from the crystal symmetries and reflects the effects of orbital degree of freedom on the magneto-electronic transport in $Sr_2IrO_4$. Note that the AMR observed in our point contacts can be as large as 14%, which is very large when compared to the crystalline AMRs reported previously in 3d transition metal alloys/oxides (0.1-0.5%) [20-22] and is also much larger than the crystalline AMRs observed to date in $Sr_2IrO_4$ [6].

It is known from the resonant x-ray [15] and neutron [20, 21] scattering experiments that $Sr_2IrO_4$ exhibits a meta-magnetic transition in an external magnetic field: the order of uncompensated magnetic moments within $IrO_2$ planes changes above the critical field $\mu_0 H_c \approx$ 200 mT. As illustrated in Fig. S1 the canting of $J_{eff} = 1/2$ moments leads to an uncompensated (residual) moment within each of $IrO_2$ planes, and these uncompensated moments can be aligned by an external magnetic field that results in a non-zero net (weakly ferromagnetic) moment at high fields. The observed point-contact MR – $R(H)$ traces in Fig. 1a – correlate very well with this transition and previously observed MRs in bulk $Sr_2IrO_4$ samples [14]. We thus attribute the observed variations in $R(H)$ to the field-induced variations in the magnetic order of $Sr_2IrO_4$ while the observed angular variations in $R(\theta)$ can be attributed to the crystalline AMR [6]. The latter is further confirmed by correlations between the observed symmetry of AMR and the crystal structure as we discuss next.

The intriguing magnetic field dependence of the AMR symmetry obtained in this work indicates yet unexplored entanglements of crystal structure, magnetic order and electron transport in this canted AFM Mott insulator. The range of external magnetic fields we applied in this study (up to ~ 0.3 T) covers the field-induced variations in the magnetic order of $IrO_2$ planes, but is too small (compared with the exchange field) to significantly alter the underlying AFM order of the $Sr_2IrO_4$ crystal. The observed AMR transition from the four- to two-fold symmetry (see Fig. 3) occurs over the same field range suggesting its possible relationship with the magnetic transformations in $IrO_2$ planes. As was pointed out in previous studies of the magnetic order in $Sr_2IrO_4$ [23-25], the magnetic moments tend to follow octahedral-site rotation because of a strong spin-orbit coupling and therefore a strong single ion anisotropy in $Sr_2IrO_4$. Since the electronic properties of $Sr_2IrO_4$ are also known to be sensitive to lattice distortions [14, 19], it is possible that the observed magnetic field dependence of AMR symmetry in our study can be associated with lattice distortions that originate from the magnetoelastic effect and spin-orbit coupling.

To explain the different AMR symmetries observed at low (four-fold) and high (two-fold) fields, we note that our $Sr_2IrO_4$ sample has a tetragonal crystallographic structure as verified by X-ray diffraction (XRD). Based on the XRD data (not shown), we found that the minima of the four-fold AMR pattern observed at low fields, where the magnetoelastic effect is small, correspond to the applied field oriented along the Ir-O bonds, assuming small octahedral distortions. Note that the four-fold symmetry of AMR is most readily observed around $\mu_0 H$ ~ 40 mT where we also see the maximum in $R(H)$ traces (Fig. 1a). At this 'coercive' field, we expect to have no net magnetization in the crystal and the measured AMR to reflect the intrinsic electronic properties of $Sr_2IrO_4$. This correlation suggests that the four-fold AMR pattern observed at low fields can be associated with the intrinsic crystal structure, band structure and orbitals of 5d electrons with spin-orbit interactions.

The two-fold symmetry of AMR observed at higher fields can be tentatively associated with the uniaxial anisotropy of the canted antiferromagnetic configuration in $Sr_2IrO_4$. Although the crystal structure is tetragonal, the canted antiferromagnetic order is orthorhombic with twinning domains as confirmed by neutron scattering/diffraction [24]. The observed angular dependence of the coercive field $H^*(\theta)$ has a predominantly uniaxial character (see Fig. 2b) that is also consistent with the existence of orthorhombic magnetic structure in the a-b plane. We attribute the uniaxial symmetry of AMR observed at low fields to the effect of the applied magnetic field on the canting of AFM magnetic moments. When the field is applied along the spin axis (near the [100] a-axis) of $Sr_2IrO_4$ the canting is reduced, while a perpendicular (to the spin axis) field would promote a larger canting. Smaller canting corresponds to a 'more antiparallel' state which corresponds to a higher resistance and thus results in a two-fold symmetry. In light of the fact that the canting of magnetic moments can be locked to the distortions of octahedra, the proposed picture of AMR due to the field mediated canting is in good agreement with previously reported magnetoelastic effect on the resistivity of $Sr_2IrO_4$ [14, 19]. The uniaxial magnetic anisotropy due to this reduced symmetry may be strengthened by the applied magnetic field, which aligns the uncompensated (residual) moments of $IrO_2$ planes [15]. Note that the observed symmetry of AMR becomes mostly two-fold at a relatively low field (~ 60 mT), indicating the predominance of the anisotropy upon the breaking of AFM order between the uncompensated moments. After that, the magnetic order (dominated by the two-fold symmetry) continues to change up to a higher magnetic field of the order of the critical field where the two-fold symmetry finally stabilizes/saturates. Moreover, the lattice distortions induced by magnetoelastic coupling are expected to further enhance the uniaxial anisotropy because of the orthorhombic magnetic structure. All of the above suggests that the field-induced lattice distortions due to the magnetoelastic effect may dominate the AMR and result in the two-fold symmetry at high fields.

In summary, we observe a large (up to 14%) AMR in point contacts to a single crystal of antiferromagnetic Mott insulator $Sr_2IrO_4$. The observed AMR has an intriguing transition from four-fold to two-fold symmetry with increasing magnetic field, which provides an interesting insight into correlations between the orbital states, electronic properties and magnetic properties of this antiferromagnetic oxide. Finally, the observed large AMR effect in a purely antiferromagnetic system without interference of ferromagnetic materials supports the development of antiferromagnetic spintronics where antiferromagnets are used in place of ferromagnets. The observed AMR that originates from strong spin-orbit interactions in 5d transition metal oxides could be used in spintronics to monitor the AFM order parameter in a more technologically favorable fashion.

The authors thank H. Chen and A. H. MacDonald for valuable discussions. This work was supported in part by C-SPIN, one of six centers of STARnet, a Semiconductor Research Corporation program, sponsored by MARCO and DARPA and by NSF grants DMR-1207577 and DMR-1122603. The work at University of Kentucky was supported by NSF via grant DMR-1265162.

[1] A. S. Núñez, R. A. Duine, P. Haney, and A. H. MacDonald, *Phys. Rev. B* **73**, 214426 (2006).

[2] A. H. MacDonald, M. Tsoi, *Phil. Trans. R. Soc. A* **369**, 3098 (2011).

[3] J. Bass, A. Sharma, Z. Wei, M. Tsoi, *Journal of Magnetics* **13**(1), 1 (2008).

[4] Z. Wei *et al*., *Phys. Rev. Lett.* **98**, 116603 (2007).

[5] W. H. Meiklejohn, C. P. Bean, *Phys. Rev.* **102**, 1413 (1956).

[6] X. Marti *et al*., arXiv: 1303.4704 (2013).

[7] B. G. Park *et al*., *Nature Mater.* **10**, 347–351 (2011).

[8] X. Marti *et al*., *Phys. Rev. Lett.* **108**, 017201 (2012).

[9] B. J. Kim et al., Phys. Rev. Lett. 101, 076402 (2008).

[10] F. Wang, T. Senthil, *Phys. Rev. Lett.* **106**, 136402 (2011).

[11] D. Pesin, L. Balents, *Nature Physics* **6**, 376 (2010).

[12] Y. Okamoto *et al*., *Phys. Rev. Lett.* **99**, 137207 (2007).

[13] G. Cao *et al*., *Phys. Rev. B* **57**, 11039(R) (1998).

[14] M. Ge *et al*., *Phys. Rev. B* **84**, 100402 (R) (2011).

[15] B. J Kim *et al*., *Science* **323**, 1329 (2009).

[16] G. Cao *et al*., *Phys. Rev. Lett.* **78**, 1751 (1997)

[17] A. G. M. Jansen, A.P. van Gelder, P. Wyder, *J. Phys. C* **13**, 6073 (1980).

[18] M. Tsoi *et al*. *Phys. Rev. Lett.* **80**, 4281(1998).

[19] L. Miao, H. Xu, Z. Q. Mao, *Phys. Rev. B* **89**, 035109 (2014).

[20] A. A. Rushforth *et al.*, *Phys. Rev. Lett.* **99**, 147207 (2007).

[21] P. N. Hai *et al. Appl. Phys. Lett.* **100**, 262409 (2012).

[22] Z. Ding *et al.*, *J. Appl. Phys.* **113**, 17B103 (2013).

[23] F. Ye *et al*., *Phys. Rev. B* **87**, 140406(R) (2013).

[24] C. Dhital *et al*., *Phys. Rev. B* **87**, 144405 (2013).

[25] S. Boseggia *et al*., *J. Phys.: Condens. Matter* **25**, 422202 (2013).

**Figure Captions:**

**Figure 1**. (a) Point-contact magnetoresistance (MR) for two point contacts of different size: PC1 with $a \approx 3$ μm (bottom) and PC2 with 1 μm (top). The black (grey) traces show the up- (down-) MR sweeps. The insert shows a schematic of our experiment: a point contact of size $a$ between a sharpened Cu tip (top grey) and the $Sr_2IrO_4$ crystal (dark) is used to inject an electrical current into the crystal; flowing primarily along the [001] c-axis into a Cu back electrode (bottom grey); magnetic field is applied in the a-b plane at an angle $\theta$ between the field and the [100] a-axis. (b) Angular dependence $R(\theta)$ of PC1 contact resistance for different magnetic fields (from top to bottom $\mu_0 H$ = 40, 60, 80, 100, 150, 200 mT). (c) Polar plots of the normalized AMR at $\mu_0 H = 40$ mT (black) and 270 mT (grey). All MR measurements were done at $T = 77$ K.

**Figure 2**. (a) Variation of the AMR magnitude with the applied magnetic field for PC1 (grey) and PC2 (black). (b) Angular dependence of the 'coercive field' $H^*(\theta)$ (black) for PC1. For comparison the grey trace shows the low field AMR data (top curve in Fig. 1a).

**Figure 3**. 2D grey-scale plots of the normalized AMR vs magnetic field for PC1 (a) and PC2 (b). Lighter color indicates higher resistance (black=0; white=1). The transitions from the four-fold (at low fields) to two-fold AMR symmetry can be seen in both plots but at slightly different fields: ~ 40 mT for larger PC1 (a) and ~ 60 mT for smaller PC2 (b).

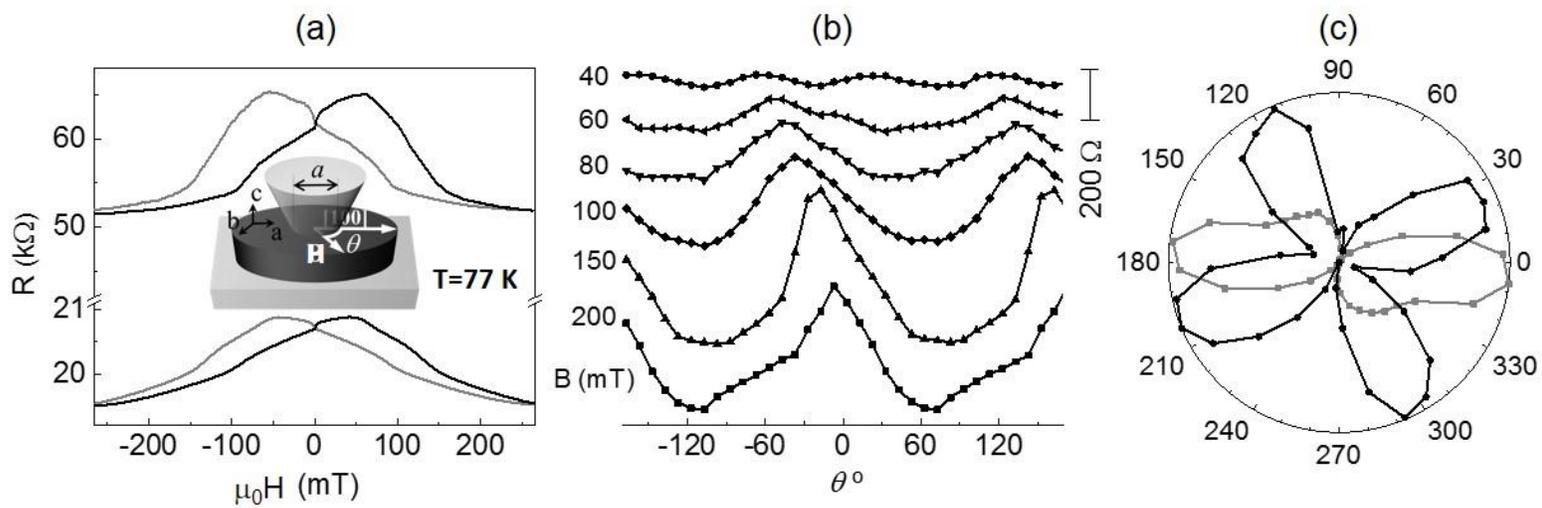

**Fig. 1**

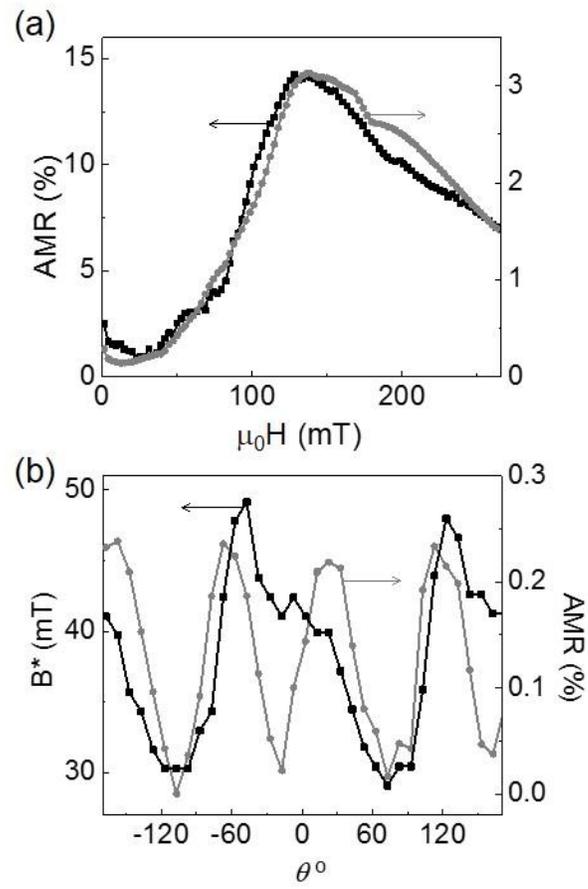

**Fig. 2**

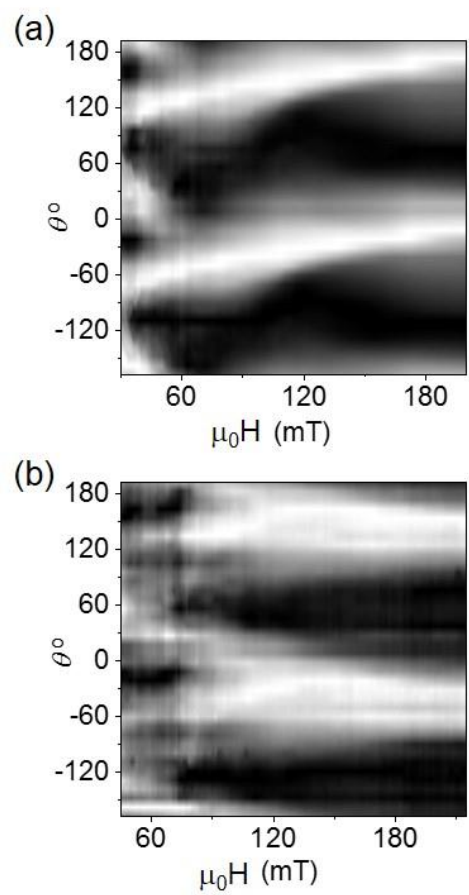

**Fig. 3**